\documentclass[amssymb,aps,prb,a4paper,twocolumn,superscriptaddress]{revtex4}
\usepackage{amsmath}
\usepackage{amsfonts}%
\usepackage{amssymb}
\usepackage{graphicx}
\usepackage{rotating}

\begin{document}

\title{A hybrid method coupling fluctuating
hydrodynamics and molecular dynamics for the simulation of macromolecules}
\author{G. Giupponi}
\email[e-mail: ]{g.giupponi@ucl.ac.uk}
\affiliation{Centre for Computational Science, Department of Chemistry, University
College London 20 Gordon street, London WC1H 0AJ, United Kingdom}
\author{G. De Fabritiis}
\email[e-mail: ]{gdefabritiis@imim.es}
\affiliation{Computational Biochemistry and Biophysics Lab (GRIB-IMIM/UPF), Barcelona Biomedical Research Park (PRBB), C/ Doctor Aiguader 88, 08003
Barcelona, Spain}
\author{Peter V. Coveney}
\email[e-mail: ]{p.v.coveney@ucl.ac.uk}
\affiliation{Centre for Computational Science, Department of Chemistry, University
College London 20 Gordon street, London WC1H 0AJ, United Kingdom}

\begin{abstract} 

We present a hybrid computational method for simulating the dynamics of
macromolecules in solution which couples a mesoscale solver for the fluctuating
hydrodynamics (FH) equations with molecular dynamics  to describe the
macromolecule.  The two models  interact through a dissipative Stokesian term
first introduced by Ahlrichs and D\"unweg [J. Chem. Phys. {\bf 111}, 8225
(1999)]. We show that our method correctly captures the static and dynamical
properties of polymer chains as predicted by the Zimm model. In particular, we
show that the static conformations are best described when the ratio
$\frac{\sigma}{b}=0.6$, where $\sigma$ is the Lennard-Jones length parameter
and $b$ is the monomer bond length. We also find that the decay of the Rouse
modes' autocorrelation function is better described with an analytical
correction suggested by Ahlrichs and D\"unweg. Our FH solver permits us to
treat the fluid equation of state and transport parameters as direct simulation
parameters. The expected independence of the chain dynamics on various choices
of fluid equation of state and bulk viscosity is recovered, while excellent
agreement is found for the temperature and shear viscosity dependence of centre
of mass diffusion between simulation results and predictions of the Zimm model.
We find that Zimm model approximations start to fail when the Schmidt number
$Sc \lessapprox 30$. Finally,  we investigate the importance of fluid
fluctuations and show that using the preaveraged approximation for the
hydrodynamic tensor leads to around $3$\% error in the diffusion coefficient
for a polymer chain when the fluid discretization size is greater than $50\AA$. 

\end{abstract}
\maketitle

\section{Introduction}
\label{intro}

The static and dynamical properties of macromolecules in solution are important
for a variety of systems, commonly referred to as soft-matter, such as
polymers, colloids, self-assembled structures, etc. Such systems are heavily
affected by the dynamical behaviour of their microscopic
components\cite{LarsonBook}, but not necessarily by their chemical details.
This allows for a coarse-grained description of solute molecules which ignores
molecular specificity\cite{DoiBook}. Indeed, the coarse grained modelling of
macromolecules in solution can be a challenge, as time and space scales
involved in such processes can easily span a range between tens of femtoseconds
and microseconds, i.e. eight orders of magnitude. In addition, coarse-grained
dilute systems can be computational overwhelming with the greater part of the
computational effort spent resolving the solvent rather than the solute, as the
ratio of the number of solvent particles to the number of solute particles can
be very large. Nonetheless solvent molecules, in which macromolecules are
embedded, must be accounted for since long range correlations produced by
hydrodynamic interactions contribute significantly to the dynamical behaviour
of macromolecules.  In these cases, it is not possible to approximate the
solvent by a thermal bath (Langevin dynamics) that disregards momentum
transport.  

Recently, various methods have been developed in order to avoid the explicit
inclusion of solvent particles while still retaining the hydrodynamic
interactions.  This is achieved by coupling a fluctuating hydrodynamics (FH)
solver to the macromolecular dynamics, which is solved using standard molecular
dynamics (MD) techniques. We refer to these approaches as hybrid methods since
the resulting dynamics of the system is taken care of by two different solvers.
Ladd\cite{Ladd93} was the first to include hydrodynamic effects in solid-fluid
suspensions while developing a method to couple colloidal particles with a
fluctuating lattice Boltzmann solver. Sharma and Patanakar\cite{Sharma04} later
extended Ladd's approach by including fluctuating Navier-Stokes equations in the
study of Brownian motion. These authors report good agreement between simulation
results and analytical and experimental data\cite{Ladd93, Sharma04}. A
second method was introduced by Ahlrichs and D\"unweg\cite{Dunweg99} to study
polymer dynamics.  They coupled the polymer chain dynamics to a fluctuating
lattice Boltzmann solver\cite{SucciBook} using a dissipative term
\cite{Ahlrichs98}.  They were able to correctly capture the effects of
hydrodynamic interactions on the chain dynamics, while Usta et
al.\cite{Usta05}, using the same method, also studied the diffusion of polymer
chains in a channel. A third method uses stochastic rotation dynamics (SRD),
also known as multiparticle-collision dynamics (MPCD) as the FH solver. The
chain dynamics is coupled to the solvent by including the monomers in the chain
in the collision step\cite{Mussawisade05}.  This method has been used to study
many different systems, including polymer \cite{Mussawisade05} and
colloid\cite{Padding06} dynamics, block copolymers\cite{Lee06} and shear
thinning\cite{Ryder06}. Both methods report a considerable computational speed
up when compared to particle-based, explicit solvent MD simulations but still
retain the effect of hydrodynamics interactions.

In this work, we present an implicit solvent method based on the approach of
Ahlrichs and D\"unweg\cite{Dunweg99} but using a recently developed
solver\cite{Gianni06} for the fluctuating hydrodynamics (FH) equations. This
solver provides an accurate description of the solvent from both hydrodynamic
and thermodynamic points of view\cite{GianniPRL}. In addition, the fluid
properties such as the equation of state and transport parameters such as shear
and bulk coefficients provide direct numerical input in this model, allowing us
to inspect the dependence of macromolecular dynamics on fluid parameters.

In the following,  we give an exhaustive description of the model.  In
particular, we illustrate the FH equations; we describe in detail the
coarse-graining procedure used to model fully flexible polymers and how
simulation parameters are directly calculated from fully atomistic simulation
force fields.  In section \ref{results} we report the static and dynamical
properties of fully flexible polymers, then we illustrate the effects of
different solvent parameters on the chain dynamics.  Finally we investigate the
role of fluctuations, providing an estimated length-scale beyond which they can
be ignored when studying macromolecular diffusion. Section \ref{conclusions}
contains our conclusions and some directions for future work.

\section{The model}
\label{model}
We integrate the fluctuating hydrodynamics (FH) equations \cite{LandauBook} for
an athermal compressible fluid over a cubic lattice using a finite volume
discretization method as proposed by De Fabritiis et al.  \cite{Gianni06} corresponding to the fluctuating hydrodynamics equations
\begin{align}
  \partial_{t}\rho & =-\partial_{\beta}g_{\beta},\nonumber \\
  \partial_{t}g_{\alpha} & = -\partial_{\beta} \left( g_{\beta} v_{\alpha}+
\Pi_{\alpha\beta}+\widetilde{\Pi}_{\alpha\beta} \right)  ,\label{fheq}%
\end{align}
where $\rho({\bf r},t)$ is the density field of the fluid, $v_{\alpha}({\bf
r},t)$ is the continuous velocity field in the component $\alpha$ and
$g_{\beta}({\bf r},t)=\rho ({\bf r},t)v_{\beta}({\bf r},t)$ is the momentum
field.  $\Pi_{\alpha\beta}({\bf r},t)$ and $\widetilde{\Pi}_{\alpha\beta}({\bf
r},t)$ are respectively the average (Navier-Stokes) and fluctuating stress
tensor fields. The average stress tensor is defined as
\begin{equation}
\mathbf{\Pi}=(p+\pi)\mathbf{1}+\overline{\mathbf{\Pi}}, 
\label{str_tens}
\end{equation}
where $p$ is the
thermodynamic pressure given by the equation of state for the fluid, $\pi = -
\zeta \partial_{\gamma} v_{\gamma}$ and $\overline{\Pi}_{\alpha\beta} =- \eta
\left( \partial_{\alpha}v_{\beta} + \partial_{\beta}v_{\alpha} - 2
D^{-1}\partial_{\gamma}v_{\gamma} \delta_{\alpha\beta} \right)$ where $\eta$
and $\zeta$ are the shear and bulk viscosity respectively and $D$ is the
spatial dimensionality. In this method it is possible to impose an equation of
state for the fluid while fluctuations are included by adding a stochastic term
to the pressure tensor, characterized by the tensor
$\widetilde{\Pi}_{\alpha\beta}$ (see Ref.\cite{LandauBook}) which is a random
Gaussian matrix with zero mean and correlations given by

\begin{align}
 \langle\widetilde{\Pi}_{\alpha\beta}(\vec{r}_1,t_1)
\widetilde{\Pi}_{\delta\gamma}(\vec{r}_2,t_2)\rangle =& 2 k_B T C_{\alpha
  \beta \gamma \delta} \delta(t_1-t_2) \delta(\vec{r}_1-\vec{r}_2),
\label{correq}
\end{align}
where $C_{\alpha \beta \gamma \delta}=\left[ \eta
  (\delta_{\alpha\delta}\delta_{\beta\gamma}
  +\delta_{\alpha\gamma}\delta_{\beta\delta}+ (\zeta-\frac{2}{D}\eta)
  \delta_{\alpha\beta}\delta_{\delta\gamma} \right]$, $k_B$ is the Boltzmann
constant and $T$ is the temperature. Note that this spatial delta-correlated
quantity, in the discrete limit of a small volume and small time
interval, can be
rewritten as
\begin{align}
  \langle\widetilde{\Pi}_{\alpha\beta}(\vec{r}_1,t_1)
  \widetilde{\Pi}_{\delta\gamma}(\vec{r}_2,t_2)\rangle \approx& \frac{2 k_B
    T}{\Delta t \Delta V}C_{\alpha \beta \gamma \delta},
\label{discretecorr}
\end{align}
where $\Delta V$ is the small volume element of fluid and $\Delta
t$ is the time step.

In the fluctuating hydrodynamics description each cell  is considered to be
delta-correlated with any other ones in space and time as shown in Eq.
(\ref{correq}).  The magnitude of the fluctuations is then determined by
temperature and viscosity; it is inversely proportional to the volume of the
discretization cell $a^3$ and the time lapse $\Delta t$, where $a$ is the
chosen lattice spacing (Eq. \ref{discretecorr}). For volumes close to the
molecular scale one would expect that fluctuations are not delta-correlated in
a molecular system, limiting the minimum size over which a discrete fluctuating
hydrodynamics description is valid. In practical terms, a box of size
$15^3-20^3 {\AA}^3$ has already proved to give good agreement between molecular
dynamics and continuum descriptions \cite{GianniPRL}.  The magnitude of the
fluctuations is also important in determining the integration step for the
stochastic differential equations.  The FH model currently uses a simple
stochastic Euler scheme to integrate the equations of motion. The time step
also depends on the mass of the fluid cell.  If the scales are small enough for
the fluctuations to appear, a good estimate of the correct time step can be
produced by a scaling analysis, using the thermal energy, mass and typical size
as $[t]^{-1} = \sqrt{(k_BT/M)}/a$, where $M$ is the mass of a cell of fluid.
Because fluctuations must satisfy the fluctuation-dissipation theorem, the
equilibrium kinetic temperature  is a good property with which to check if the
size of the timestep is appropriate.

In the FH model, transport coefficients such as shear and bulk viscosities,
which heavily influence macromolecular dynamics, are input parameters. In this
paper, we report results obtained by imposing periodic boundary conditions
using parameters specific to liquid water. All quantities are reported in $[l]
= \AA$, [m] = $g/mol$, [T] = Kelvin and [E] = $Kcal/mol$ unless otherwise
stated. We note that the time unit is a derived quantity and is equal to $48.8$
femtoseconds. For water at $T = 300 K$ and $p=1$ atm, we set shear and bulk
viscosities to $\eta = 2.6$, $\zeta = 6.2$ respectively\cite{Gianni06}. In
addition, FH enables boundary conditions such as Couette and Poiseulle flows to
be readily implemented, which are necessary in order to study rheological
properties of complex fluids such as mixtures of fluids and macromolecules
(i.e. polymers in a solvent)\cite{Gianni06}. 

We focus here on the dynamics of fully flexible polymers. We model a fully
flexible polymer as a set of Lennard-Jones (LJ) monomers  
\begin{equation} 
V_{nb}(r) = \left\{ \begin{array}{rl}
4\epsilon[(\frac{\sigma_{lj}}{r})^{12} - (\frac{\sigma_{lj}}{r})^{6} +
\frac{1}{4}]  & r \le r_{cut}; \\ 
0 & r > r_{cut}
\end{array}\right .
\label{LJeq}
\end{equation} 
where $\sigma_{lj}$ and $\epsilon$ are respectively LJ length and energy units.
$V_{LJ}(r)$ is truncated at a cutoff radius $r_{cut}$,
depending on the solvent quality modelled, and is shifted by a factor 1/4 to
avoid an energy discontinuity when $r_{cut}$ is set equal to the potential
minimum\cite{AllenBook}. We set $r_{cut}$ to $V_{nb}(r)$ minimum ($r_{cut} =
2^{1/6}\sigma$) when mimicking good solvent conditions, i.e. the potential is purely
repulsive, whereas an attractive tail is added when simulating poor solvent
conditions.  

A spring potential is introduced to model chain connectivity 
\begin{equation} 
V_b(r) = K_b(r - b)^2, 
\label{SPRINGeq}
\end{equation} 
where the spring constant $K_b=0.8 Kcal/mol/{\AA}^2$ is chosen large enough to limit the
fluctuations in the polymer radius of gyration to less than 20 per cent.  
The bond length $b=\frac{5}{3}\sigma$ is chosen to match the theoretical 
static scaling exponent, $\nu = 0.588$ (see below).

Due to the hybrid nature of our model, it is convenient to use the same units
for MD and FH.  This requires a clear understanding of the coarse graining
procedure that allows macromolecules to be modelled as a collection of
LJ-interacting beads connected by springs\cite{KremerGrest90}. We use the
chemical formula of a standard polymer, namely polyethylene, to coarse-grain
$3-4$ repeated units as a single bead, i.e. a ``monomer". Using atomistic MD
forcefields as a reference, we estimate an excluded volume parameter for such a
monomer of $\sigma = 15\AA$, and $m = 50-100$ a.m.u. Similar estimates can be
found in the literature\cite{Pitard99}. We set $\epsilon = 1.2k_bT$ where $k_bT =
0.6 Kcal/mol$ at 300 K. The equations of motion are integrated using the
velocity Verlet algorithm. It is important to stress that the timestep used to
integrate the fluctuating hydrodynamics equations can differ from the MD timestep.
Here, however, we set the two integration timesteps to be $\Delta t = 10$ femtoseconds.

Finally, the coupling between MD and fluid dynamics is implemented following
Ahlrichs and D\"unweg\cite{Ahlrichs98}. They model a monomer as a point-like
object which interacts with the fluid via a friction term to represent the
viscous force $\vec{F}_i$ exerted by the fluid on monomer $i$ 
\begin{equation} 
\vec{F}_i = -\zeta_b[\vec{v}_i(\vec{r}) - \vec{u}_{f}(\vec{r})] + \vec{f} , 
\label{COUPLeq}
\end{equation} where $\zeta_b$ is the ``bare" friction coefficient,
$\vec{v}_i(\vec{r})$ and $\vec{u}_f(\vec{r})$ are respectively the velocity of
the monomer and the fluid at position $\vec{r}$, and $\vec{f}$ is a stochastic
force \cite{Ahlrichs98}. The fluid velocity at position $\vec{r}$, 
$\vec{u}_f(\vec{r})$, is calculated using a linear interpolation of grid point
velocities\cite{Dunweg99}. The same interpolation scheme is used to transfer the
force from the monomer to the fluid, thus ensuring conservation of total
momentum in the system.

As noted before, a crucial parameter in the model is the grid spacing size $a$
which determines the discretization volume. This is very important when trying
to coarse grain a physical system, as $a$ is the minimum scale at which
hydrodynamic interactions can be resolved. Moreover, $a$ influences the effective
diffusion of a monomer coupled to the fluid, as the effective monomer friction
is\cite{Dunweg99} 
\begin{equation} 
\frac{1}{\zeta_{eff}} = \frac{1}{\zeta_{bare}} + \frac{1}{g\eta a} , 
\label{Geq} 
\end{equation} 
where the
factor $g$ takes into account the lattice geometry\cite{Dunweg99}.  In other
words, the effective friction is the sum of a term related to the Brownian
motion due to uncorrelated collisions with fluid particles and a term which
takes into account the hydrodynamic velocity field.  One aim of the present paper
is to measure our factor $g$ once and for all.  Eq.\ref{Geq} can also be
considered as a first consistency check for this model.  
\begin{center}

(figure 1: g calculation )

\label{f1}
\end{center}
We calculate the mean-square displacement $<(\vec{r}(t) - \vec{r}(0))^2>$ of a
monomer with different bare frictions $\zeta_b$. The monomer is embedded in a
$500\AA \times 500\AA \times 500\AA$ fluid box with periodic boundary
conditions. We set the temperature $T=300K$ and pressure $p = 1$ atm and derive the
effective friction coefficient $\zeta_{eff}$ via the relation $<(\vec{r}(t) -
\vec{r}(0))^2> = \frac{k_bT}{\zeta_{eff}}t$.  We plot in fig.~1 the simulation
results for different viscosities $\eta$ and bare coefficients
$\zeta_{bare}$ (see caption for details). In the inset, we show results
using different lattice sizes $a$. The agreement with eq.~(\ref{Geq}) is
excellent, leading to a value $g=45.5$ which is consistent with the result of
Ahlrichs and D\"unweg\cite{Dunweg99} and Usta et al.\cite{Usta05}, who couple
the dynamics of a polymer to a lattice Boltzmann solver. We also confirm that
the effective diffusion does not depend on the monomer
mass\cite{Giupponi06dsfd}. This is consistent for the present case,
where the monomer dynamics is primarily influenced by viscous forces. In
addition, it is important to note that a central assumption in the
Rouse-Zimm theory of polymer dynamics is that inertial effects are
negligible\cite{DoiBook}. Indeed, Rouse-Zimm models assume the overdamped
Smoluchoski equation for the dynamics of a monomer in a solvent.  Therefore,
our findings demonstrate that this approximation actually holds in the present case.

\section{Results}
We begin by studying the static and dynamical properties of fully flexible
chains.  Our model correctly captures the radius of gyration $(R_g)$ dependence
on the number of monomers in a chain $(N)$ and the scaling law for the monomer
diffusion as predicted by theory\cite{DoiBook}. We also provide numerical
evidence for an improved formula proposed by Ahlrichs and
D\"unweg\cite{Dunweg99} for the relaxation of Rouse modes. In the following
subsection we investigate the dependence of macromolecular dynamics on the
different solvent parameters.  Finally, we assess the importance of fluid
fluctuations by calculating the chain diffusion coefficient for different
system sizes. 
\label{results}
\subsection{Static and dynamic properties of polymer chains}
\label{chain_prop}

{\it Chain statics} - The radius of gyration $R_g$ of fully flexible polymer
chains scales with the number of monomers N as a power law\cite{DoiBook},
$<R_g^2>^{\frac{1}{2}} \sim N^{\nu}$. The exponent $\nu$ is
crucial for the description of both static and dynamical properties of
polymer chains and is equal to 0.588.
In fig.~2, we plot the spherically averaged static structure factor $S(q)$ for
chains comprised of $N=20,40,50,100$ monomers
\begin{equation}
S(q) = \frac{1}{N}\sum_{j=1}^{N}\sum_{k=1}^{N}<\frac{\sin \left( q\left| 
\vec{r}_{j}-\vec{r}_{k}\right| \right) }{q\left| \vec{r}_{j}-
\vec{r}_{k}\right|}>, 
\end{equation} 
where $\vec{r}_{j}$ is the position vector of the $j$-th monomer, $q$ is the
magnitude of the scattering wave vector and $<\cdots>$ is an ensemble average .
$S(q) \sim q^{-1/\nu}$ in the range $(2\pi/<R_g^2>^{1/2}) \ll q \ll (2\pi /
\sigma) $\cite{DoiBook}, allowing us to calculate $\nu$ by fitting the data from
only one simulation. In addition, it is more efficient to collect independent chain
configurations as the autocorrelation times of $S(q)$ in the range of the fit are
shorter than the radius of gyration autocorrelation time for given N. We
therefore determine $\nu$ by calculating $S(q)$ using chains with a varying
number of monomers
$N=20,40,50,100$. Our results are shown in fig.2. 
\begin{center}

(figure 2: S(q) static )

\end{center}
We obtain $\nu = 0.581 \pm 0.005$ for our chosen set of parameters, $b = 15\AA$
and $\frac{\sigma}{b}=0.6$, which agrees well with the theoretical value\cite{DoiBook}. We
note that for the $N=100$ chain, $\nu =0.588$. The particular choice of the ratio
$\frac{\sigma}{b}$ is explained by the systematic dependence of the calculated
$\nu$ on the ratio $\frac{\sigma}{b}$, as shown in the inset of fig.2 for an
$N=50$ chain.  Increasing $\frac{\sigma}{b}$ from 0.2 to 1, we obtain a
decreasing slope and, therefore, an increasing value of the scaling exponent,
$\nu \in [0.512, 0.628]$. This is due to the fact that the theoretical result
for $\nu$ is obtained in the thermodynamic limit in which $N\rightarrow \infty$.
Therefore, our choice is an empirical way to compensate for the finite length
of the modelled chain. We point out that different simulation methods such as
dissipative particles dynamics\cite{Syme05} (DPD) and stochastic rotational
dynamics\cite{Mussawisade05} (SRD) have obtained $\nu \sim 0.62$ using the more
standard $\sigma = b$ ratio, which is very similar to our own result for
$\sigma = b$. Ahlrichs and D\"unweg\cite{Dunweg99} also obtained $\nu = 0.62$,
and referred to this slightly different estimate of $\nu$ as one of the main
problems to be addressed in order to improve their model: they believed this
discrepancy to be the main source of error in their simulations.  Finally, as
static properties do not depend on properties of the bath, we also checked the
dependence of the calculated value of $\nu$ on the ratio $\frac{\sigma}{b}$
using Brownian dynamics, obtaining the same qualitative results.

{\it Chain dynamics} - When calculating dynamical properties, care has to be
taken as the boundary conditions used can in general affect the simulation
results. We use periodic boundary conditions and therefore we must be able to
control the effects of unrealistic interactions of a chain with its periodic
replicas. As hydrodynamic interactions decay slowly as $r^{-1}$, these effects
can be very important, being of order $L^{-1}$, where $L$ is the box length.
Consequently, when inspecting the chain dynamics, the ratio $\frac{R_g}{L}$
should be made as small as possible. For our simulations, we used $\frac{R_g}{L}
\sim \frac{1}{5}$, in line with other hybrid models\cite{Dunweg99,
Mussawisade05, Usta05, Lee06}. 

An additional issue to be settled when choosing simulation parameters is the
influence on the dynamics of the ratio of the fluid discretization size $a$ to
the distance between connected monomers $b$. Ahlrichs and
D\"unweg\cite{Dunweg99} showed that the decay of the normalized dynamic
structure factor for a polymer chain is 20\%-25\% slower when $a=2b$ compared
to the case $a=b$, i.e. hydrodynamic effects are smaller as $\frac{a}{b}$
increases. This is a consequence of a more coarse-grained resolution of the
velocity flow field, and corresponds to the fact that on lattice grid the
hydrodynamic modes for $\tilde{k}_a \ge \frac{2\pi}{a}$ are cut off. However,
depending on the coarse graining scheme, hydrodynamic interactions might then not
extend down to the monomer scale. In addition, a fine-grained resolution of the
velocity field might not be necessary, especially when dealing with long
chains. In fact, Usta et al.\cite{Usta05} found that the diffusion coefficients
for $N>128$ chains are indistinguishable for the $a=b$ and $a=2b$ cases. We
also found congruent results\cite{Giupponi06dsfd,Giannieme06} for the diffusion
coefficients of $N>50$ chains and the collapse time of an $N=300$ polymer in
poor solvent. Our findings confirm the suggestion made by Usta et al. of a
$\frac{R_g}{a} > 5$ ratio being sufficient to extract diffusion coefficients.
In these simulations, the conservative choice $a= \frac{4b}{3}$ is used, which we
have found to be a good compromize between the requirements of capturing
hydrodynamics effects and the minimization of simulation time.  

\begin{center}

(figure 3: CoM diff N=50)

\end{center}

We plot in fig. 3 the mean square displacement (MSD) of the chain centre of
mass $ MSD_{CM}=<(r_{CM}(t) - r_{CM}(0))^2>$ for an $N=50$ chain in water as a
function of time $t$. From the relation $ <(r_{CM}(t) - r_{CM}(0))^2> = 6Dt$,
where D is the diffusion coefficient of the chain, we obtain  a value
$D_{FH}=0.026 \AA^2/ps$ from the data in fig.3.

\begin{center}

(figure 4: central mon diffusion Zimm wins over Rouse)

\end{center}
Our model correctly captures the effects of hydrodynamic interactions on the
dynamics of the chain. In fig.4 we plot $MSD_{mm}$, the mean square
displacement of the monomer in the middle of an $N=50$ chain in water. We
choose the central monomer in order to avoid end effects. The main theoretical
results on chain dynamics are contained within the Rouse and Zimm
theories\cite{DoiBook}. The Zimm model includes hydrodynamic interactions,
which are completely neglected in the Rouse model. However, both theories
predict that the monomer mean square displacement scales as $t^{\alpha}$, where
$\alpha$ is equal to $\frac{2}{3}$ or $0.54$ respectively. Our results,
obtained by fitting our data to a power-law
curve (shown in fig.4), indicate a $t^{0.658}$ scaling for the monomer
mean-square displacement. Therefore our model appropriately describes the
importance of hydrodynamic interactions for the dynamics of a chain in a dilute
solution. It is interesting to note that Zimm-Rouse theories use a standard
Oseen tensor derived on the basis of non-fluctuating, incompressible
Navier-Stokes equations\cite{DoiBook}, which is different from FH
eq.(\ref{fheq}) used in our model.  However, our simulations show that the
theoretical scaling factor still holds, at least for the coarse-graining
procedure and solvent chosen here.
\begin{center}

(figure 5: Chis decays, dunweg refinement)

\end{center}
The intramolecular dynamics is usually described in terms of Rouse modes
$\chi_p$ 
\begin{equation} 
\vec{\chi}_p = N^{-1}\sum_{n=1}^N \vec{r}_n cos[\frac{p\pi}{N}(n - 1/2)], 
\end{equation} 
where $p = 1,2,\hdots$ The Zimm model predicts an exponential decay for the
Rouse modes autocorrelation function 
\begin{equation} 
ACF_{\chi}^p = \frac{<\vec{\chi}_p(t)\vec{\chi}_p(0)>}{<\vec{\chi}_p^2>} = exp(-t/\tau_p),
\label{chi_decay}
\end{equation} 
with $\tau_p$ scaling as $p^{3\nu} \sim p^{1.77}$. In their
work, Ahlrichs and Dunweg suggested an analytical improvement to the Zimm model
($\tau_p \sim p^{1.77}r(p)$, see appendix A in ref.\cite{Dunweg99}) and found
evidence for a better agreement with simulation results. Their conclusion was
confirmed by Polson and Gallant using an explicit solvent coarse-grained MD
simulation\cite{Polson06}, but Mussawisade et al.\cite{Mussawisade05} observed
no deviation from the Zimm result by coupling stochastic rotation dynamics to
the polymer chain dynamics.  In fig.5, we plot the first five $ACF_{\chi}^p, p
= 1\hdots5$ vs. time. We compare the theoretical Zimm prediction with the
proposed analytical solution by rescaling the $x$-axes by a factor depending on
the proposed $\tau_p$ scaling formula (see caption of fig.\cite{Dunweg99} for details). Our results
show that the $ACF_{\chi}^p$ collapse is better when the $r(p)$-correction is
included in the theory. We also note that further theoretical approximations are
required in order to recover the abovementioned scaling results for $\tau_p$,
eq.(\ref{chi_decay}). These approximations are more significant for short
chains although Mussawisade et al.\cite{Mussawisade05} claimed that different
approximations would cancel out. Our data, in line with other
authors' findings\cite{Dunweg99,Polson06}, show no evidence for such
compensation.

\subsection{Solvent effects}
\label{solvent_effects}

\begin{center}

(figure 6: Temperature effects)

\end{center}

{\it Solvent temperature} - 
It is important to make sure that a mesoscopic description of the solvent
correctly captures the effect of varying solvent temperature, as
this is an important variable when studying the dynamics of
macromolecules in solution.  Fig.~1 shows that the diffusion coefficient of a
single monomer obeys eq.~(\ref{Geq}), which is implied by the
fluctuation-dissipation theorem. Ahlrichs and Dunweg\cite{Ahlrichs98} also
showed that the fluctuating force in the coupling term, eq.~(\ref{COUPLeq}), is
required in order to explicitly satisfy the fluctuation-dissipation relation $
<V(t)V(0)> = V_r(t)$, where $<V(t)V(0)>$ is the velocity autocorrelation
function of a single monomer, and $V_r(t)$ is the velocity relaxation of an
initially kicked monomer. Usta et al.\cite{Usta05} showed that $m<V^2>/k_bT =
1$ when $\zeta_{eff}\Delta t/m < 0.04$ using the same coupling scheme,
eq.~(\ref{COUPLeq}). We verify here that our model also exhibits the correct thermal
behaviour for the diffusion of a polymer chain. The Zimm and Rouse models
predict that the diffusion coefficient for the centre of mass for a
polymer chain is proportional to the solvent temperature $T$. In fig.~6 we plot
$D_{FH}$ for simulations with $T=50,100,200,300 K$. We see that the expected
proportionality holds and we therefore conclude that the proposed
coupling between the mesoscopic solvent solver and the polymer molecular
dynamics, eq.~(\ref{COUPLeq}), adapts well to the study of dynamics of
macromolecules in solution.

\begin{center}

(figure 7: Compressibility and equation of state)

\end{center}

{\it Solvent compressibility and equation of state} -
Compressibility effects are measured by the nondimensional Mach number
$Ma=\frac{v_s}{c_s}$, which is the ratio between  the speed of the solvent or
the polymer chain $v_s$ and the velocity of propagation of density waves $c_s$.
As $Ma \sim 10^{-3}-10^{-4}$ in our simulations, it is reasonable to expect
that compressibility effects  do not affect the dynamics of the chain, i.e.
perturbations in the density field are very quickly dissipated before altering
the diffusive processes.  In order to test the dependence of the dynamics of a
chain on the bulk viscosity $\zeta$ of a fluid, we calculate the diffusion
coefficients $D_{\zeta}$ for an $N=50$ chain by performing simulations with
different $\zeta$, keeping all other parameters fixed.  In fig.~7 (inset) we
plot $\frac{D_{\zeta}}{D_{FH}}$ for $\zeta=0,3,6.2,9$. Our simulation results
indeed show a negligible dependence of the diffusion coefficients on $\zeta$.
It is important to note that this result furnishes another direct confirmation
of the robustness of Zimm theory which uses the Oseen hydrodynamic tensor
derived from the incompressible Navier-Stokes equations,  i.e. $ \zeta
\partial_{\gamma} v_{\gamma}= 0$ in eq.(\ref{str_tens}).  

The solvent equation of state, which is a closure relation dictated by
thermodynamics, rules the pressure-density relation, and therefore the
amplitude and velocity of sound waves. As we are working in a
nearly-incompressible regime, it is not  expected to influence the behaviour of
the chain. To investigate the effects of a different equation of state on the
dynamics of a polymer chain, we run a simulation for an $N=50$ chain embedded
in a fluid with argon equation of state\cite{Gianni06} but water viscosity.  In
fig.~7 we plot the results for the centre of mass mean square displacement
$MSD_{CM}$ comparing it with the same result obtained in section
\ref{chain_prop} using the equation of state for water. It is clear from fig.~7
that the effect of a different equation of state on the chain diffusion is
negligible. However, we cannot completely rule out the possibility that the
equation of state may affect the macromolecular dynamics in the incompressible
regime, as the hybrid method employed here couples transversal modes only. An
answer to this question may be provided by coupling the monomer to the fluctuating
hydrodynamic pressure tensor. We reserve this study for future work.

\begin{center}

(figure 8: Diff vs eta)

\end{center}

{\it Solvent viscosity} - 
The Zimm model predicts that the diffusion $D_{FH}$ of the centre of mass of a
polymer chain is proportional to the inverse of shear viscosity, $D_{FH} \sim
\eta^{-1}$.  We run simulations with different shear viscosities
$\eta=1,2,2.6,3.5,4,5.5,6.48$ and in fig.~8 we plot the calculated diffusion
coefficient $D_{FH}$ vs. $\eta^{-1}$. We note that linear scaling is not
produced across the entire range of shear viscosities studied. However, a good
linear fit (the continuous line in fig.~8) is obtained when the $\eta=1$
result is excluded. We explain this result by recalling the fact that the Zimm model
assumes that the fluid relaxation is much quicker than the monomer diffusion,
i.e.  the momentum transport is much faster than the mass transport.  The
non-dimensional Schmidt number $Sc=\frac{\eta}{(\rho D)}$ expresses the rate of
momentum transfer relative to the rate of mass transfer in a fluid. For
example, $Sc \sim 10^8$ for a $10nm$ sedimenting colloid\cite{Cates}, and
therefore $Sc \gg 1$ is usually assumed in macromolecular dynamics theories.
Our simulation results therefore suggest a theory breakdown 
when $\eta=1$, i.e. $Sc \lessapprox 30$ in our model.

\subsection{Fluctuations and coarse-graining}
\label{fluc}

\begin{center}

(figure 9: Diff vs fluctuations, i.e. box sizes etc)

\end{center}

As explained in section \ref{model}, the magnitude of fluctuations is
proportional to the inverse of the volume of the discretization cell, $a^3$.
We therefore expect a decreasing influence of the fluctuating term
$\tilde{T}_{\alpha \beta}$ on the solutions of eq.(\ref{fheq}) as system size
increases.  In turn, the importance of fluctuations in macromolecular dynamics
will also become less significant. We quantify this using different fluid
discretization sizes $a$, and rescaling $b, \sigma$ and $\Delta t$ accordingly.
The monomer mass $m$ is taken to scale with the lattice cell volume $a^3$. We
run two simulations for each new system, with and without the fluctuation term
$\tilde{T}_{\alpha \beta}$, calculating the diffusion coefficient for an $N=50$
polymer chain. We plot in Fig.~10 $(D_F-D_{NF})/D_F$ vs the fluid
discretization size $a$, where $D_{F(NF)}$ is the diffusion coefficient
obtained with(without) fluctuations. Our simulations confirm that the
importance of fluctuations becomes smaller as the fluid discretization size $a$
increases. The percentage difference is around 7-8\% when the grid size is
$a=20\AA$, decreasing to around 3\% when the grid size is doubled to $a=50\AA$.
Our result shows that, at least for chain diffusion, fluctuations of the
fluid do not play a significant role when using a fluid discretization size
bigger than $50 \AA$, as in this case the fluctuating term in
eq.~(\ref{COUPLeq}) becomes dominant.  This is not only important as a rule of
thumb when coarse graining such physical systems, but should also help
theoretical modelling, as it assists in quantifying the degree of approximation
involved in using the preaveraged approximation.

The parameters chosen when coarse graining a system are not only important for
assessing the relative importance of different physical processes such as
fluctuations, but also for the comparison of theoretical results with
experiments. It is clear that even using hybrid methods, it is impossible to
grasp all the physics. Indeed the approach helps to put the focus on the most interesting
aspects\cite{Cates, Padding06}. Robertson et al.\cite{Robertson06}, studying the
diffusion of linear DNA chains experimentally, found a diffusion coefficient of $D_{DNA} =
1.28\times10^{-4} \AA^2/ps$, which is more than two orders of magnitude smaller than
the diffusion coefficients obtained in this study. Although coarse-graining
procedures are known to speed up dynamics\cite{Depa05} and long-time,
experimental diffusion is lower than short-time, Kirkwood
diffusion\cite{Fixman81, Liu03}, these effects are small and clearly cannot
justify such discrepancies. In order to obtain agreement between simulation
results and experiments, the MD parameters must be adjusted to the experimental
system in question. In this case, the contour length for the shortest linear DNA chain used
in ref.\cite{Robertson06} is $2.65\mu m$, whereas in our simulations an $N=50$
chain has a contour length of $0.075\mu m$. Simulations involving a chain of
thousands of monomers would therefore be necessary to concord with experimental
contour length, but this would clearly be computationally infeasible. However,
we obtain a better result using a different, DNA-tailored CG procedure,
associating the monomer bond length $b$ to a persistence length $ l_p=40nm$,
$m=10000 a.m.u.$ These are realistic values for linear DNA macromolecules and
permit us to reach a reasonable contour length using an $N=100$ chain. With an MD
integration timestep of $\Delta t = 2ps$, the calculated diffusion coefficient
is $1.1\times10^{-3}\AA^2/ps$, which is within an order of magnitude of
published experimental results. Such a CG scheme can therefore be used as a
starting point to make quantitive predictions on more complicated processes,
such as DNA translocation. As a final point, we note that independently of the
coarse graining scheme, numerical diffusion does not affect this type of
microscopic simulation, as Reynolds numbers are very low (approximately zero)
and the fluid viscosity is dominant compared to the numerical one\cite{Gianni06}.

\section{Discussion and Conclusions} \label{conclusions}

We have presented in this paper a general hybrid model to simulate the dynamics
of macromolecules in solution. The term hybrid refers to the fact that
hydrodynamic forces are calculated by a separate fluctuating hydrodynamics solver
coupled with the macromolecular dynamics. Treating the solvent implicitly
allows for a significant saving in computational time. We estimate a speed-up
factor of 30 for the simulations performed here compared to particle-based
simulations. We use a newly developed hydrodynamic solver that integrates the
fluctuating hydrodynamics equations. Fluctuations are included according to the
Landau formalism and are thermodynamically consistent. In addition, fluid
characteristics such as transport parameters and equations of state are direct
input parameters. 

We have studied fully flexible polymers in TIP3P water at $300K$ with our
model. We show that static and dynamical properties for the polymer chain are
correctly captured. In particular, the critical exponent $\nu \sim 0.588$ is
calculated using $\sigma=0.6b$ in the LJ potential while the Zimm model
prediction for the scaling of monomeric diffusion is also obtained. This
confirms that hydrodynamic interactions are quantitatively described in this
simulations framework.  In order to provide information on the much debated
issue of the scaling behaviour of the autocorrelation functions of the Rouse modes
$ACF_{\chi}^p$, we calculate these and demonstrate that the p-dependent scaling
formula suggested by Ahlrichs and D\"unweg describes the simulation results
well.  Moreover, our model relaxes the hypothesis of a non-fluctuating,
incompressible hydrodynamic tensor assumed by the Zimm model. Moreover, as
found by other authors with different
methods\cite{Dunweg99,Mussawisade05,Usta05,Polson06}, our results show the
robustness and consistency of these hypothesis for the case of a polymer chain
in a viscous and nearly incompressible fluid such as water.

We have also investigated the effect of various solvent parameters on chain
dynamics. We found the predicted direct proportionality between the diffusion
coefficient of a polymer chain and the temperature of the solvent.  As a
consequence of working in a nearly incompressible regime and with the present coupling
scheme, altering the equation of state and bulk viscosity have a 
negligible influence on chain diffusion. We tested the predicted Zimm formula
for the dependence of chain diffusion with viscosity for a range of
viscosities. Our data show a good agreement with Zimm theory when the viscosity
$\eta$ is greater than $1$. We therefore estimate that the high Schmidt number
hypothesis starts to fail at $\eta \sim 1$, i.e. around a third of the
viscosity of water. It would be interesting to test this prediction
experimentally. We point out that our results for solvent effects should hold
for a generic macromolecule diluted in solvent.

We investigated the importance of fluctuations on the dynamics of a
chain, confirming as expected that the importance of the fluctuating term in the
Navier-Stokes equations decreases as system size increases. We suggest that a lattice size
$a = 50\AA$ is a safe estimate to restrict the error introduced by the
preaveraged approximation to a few percent. Finally, we discussed the importance of
the coarse-graining method that needs to be employed in order to achieve agreement between
simulation results and experimental observations.  We obtain a realistic
diffusion coefficient for DNA chains using a tailored coarse-graining procedure which can
be used in future work concerned with the study of real systems. We also plan
to use this model to investigate the dynamics of less theoretically well understood
systems such as dendrimers, semi-flexible polymers and very dilute many-chains
systems.

\section{Acknowledgements} 
We thank I. Pagonabarraga for very helpful discussions and R. Delgado
Buscalioni and D.M.A. Buzza for a critical reading of the manuscript.  We are
grateful to EPSRC for funding under grants GR/R67699 (RealityGrid,
http://www.realitygrid.org) and GR/S72023 (ESLEA,
http://www.eslea.uklight.ac.uk).

\pagebreak

\bigskip

\noindent Figure 1\qquad Plot of nondimensional bare (effective) monomer diffusivity
versus dimensionless viscosity. We use different viscosities ($\eta = 0.27,2.6,6.2$)
and grid sizes ($a=15,20,25\AA$) to test eq.(\ref{Geq}). The agreement with eq.(\ref{Geq})
is excellent for g = 45.5.

\bigskip

\noindent Figure 2\qquad Log-log plot of the static structure factor $S(q)$ vs. q for 
$N=20,40,50,100$ polymer chains. Inset: Detail of $S(q)$ for an $N=50$ chain with 
$\frac{\sigma}{b}= 0.2,0.4,0.5,0.6,1$.

\bigskip

\noindent Figure 3\qquad The mean square displacement vs. time for the centre
of mass of an $N=50$ chain in water.

\bigskip

\noindent Figure 4\qquad The mean square displacement vs. time for the central monomer
of an $N=50$ chain in water.

\bigskip

\noindent Figure 5\qquad Log-log plot of autocorrelation function of the Rouse modes $\vec{\chi}_p$,
$p=1\hdots5$. The x-axis in the panel above(below) is rescaled according to Zimm-model 
theory prediction(Ahlrichs and D\"unweg formula\cite{Dunweg99}). Note that a better agreement
is obtained with Ahlrichs and D\"unweg formula.

\bigskip

\noindent Figure 6\qquad Plot of diffusion coefficient for an $N=50$ polymer chain vs.
solvent temperature.  

\bigskip

\noindent Figure 7\qquad Plot of mean square distance of centre of mass for an
$N=50$ chain in a fluid with argon equation of state but water viscosity
(downward pointing triangles) and water (upward pointing triangles) equations
of state. Inset: Plot of ratio of diffusion coefficients
$\frac{D_{\zeta}}{D_{FH}}$ obtained for various values of bulk viscosity
$\zeta=0,2.6,6,9$. 

\bigskip

\noindent Figure 8\qquad Plot of diffusion coefficient $D_{FH}$ vs inverse shear viscosity,
$\frac{1}{\eta}$. Note that the predicted linear relation does not hold for $\eta=1$, i.e.
$Sc \sim 30$.

\bigskip

\noindent Figure 9\qquad Plot of $\frac{D_{F}-D_{NF}}{D_F}$ vs. grid size $a$.
($D_{F(NF)}$) refers to the diffusion coefficient calculated with or without
the fluctuating term in eq.(\ref{fheq}).

\pagebreak

\begin{figure}
\begin{center}
\rotatebox{90}{\scalebox{0.8}[0.8]{\includegraphics{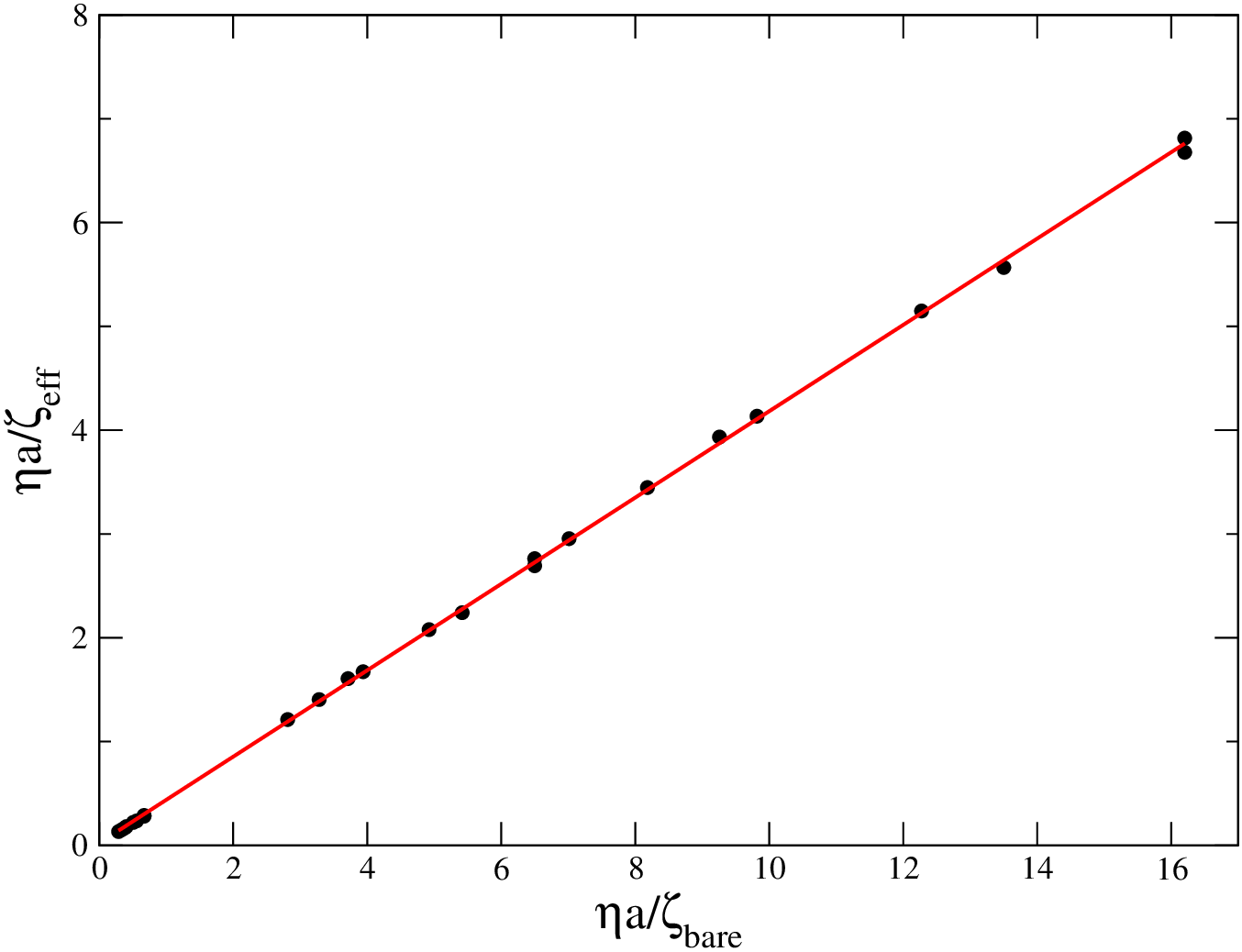}}}
\caption{}
\end{center}
\end{figure}


\bigskip

\begin{figure}
\begin{center}
\rotatebox{90}{\scalebox{0.8}[0.8]{\includegraphics{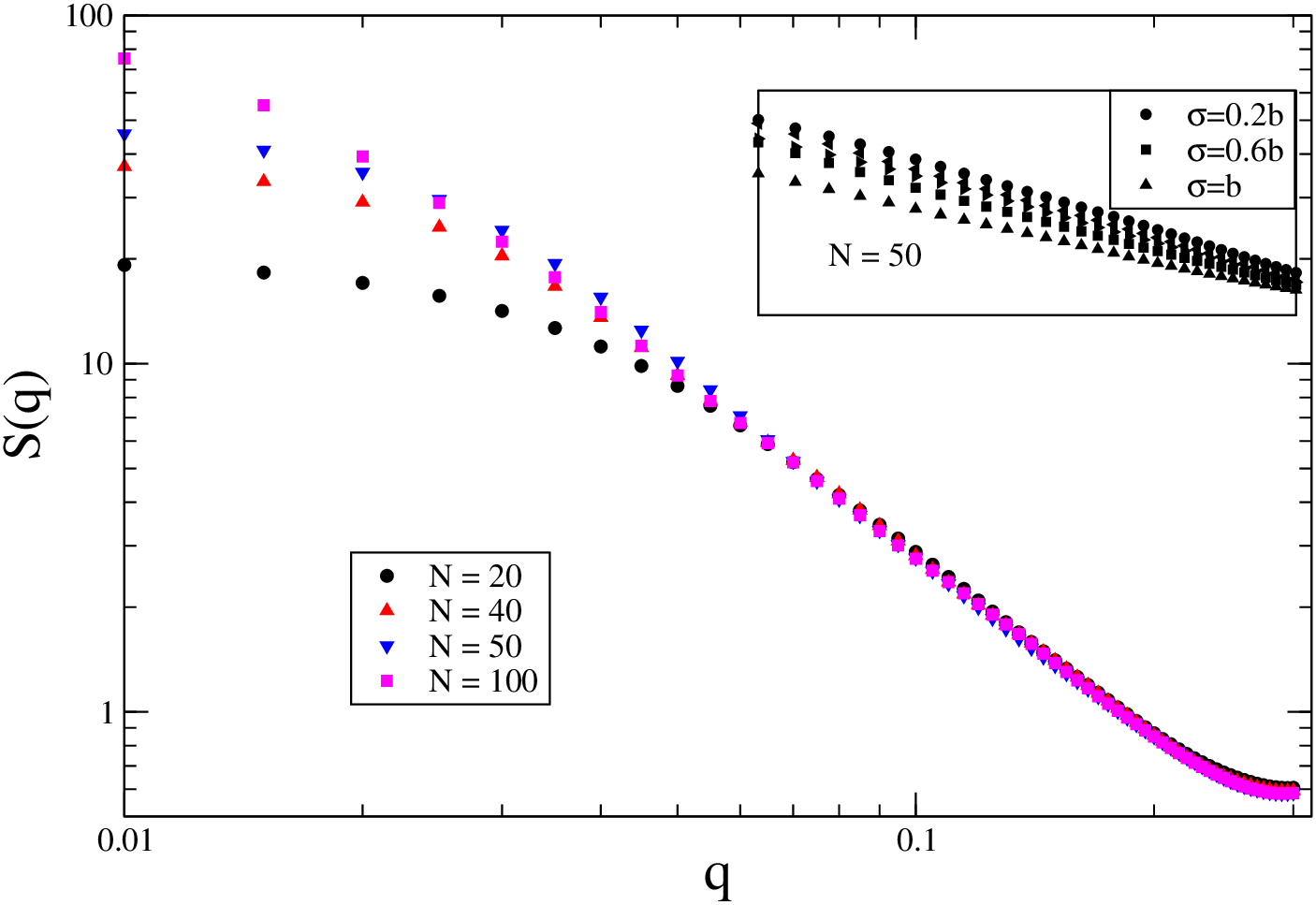}}}
\caption{}
\end{center}
\end{figure}


\bigskip

\begin{figure}
\begin{center}
\rotatebox{90}{\scalebox{0.8}[0.8]{\includegraphics{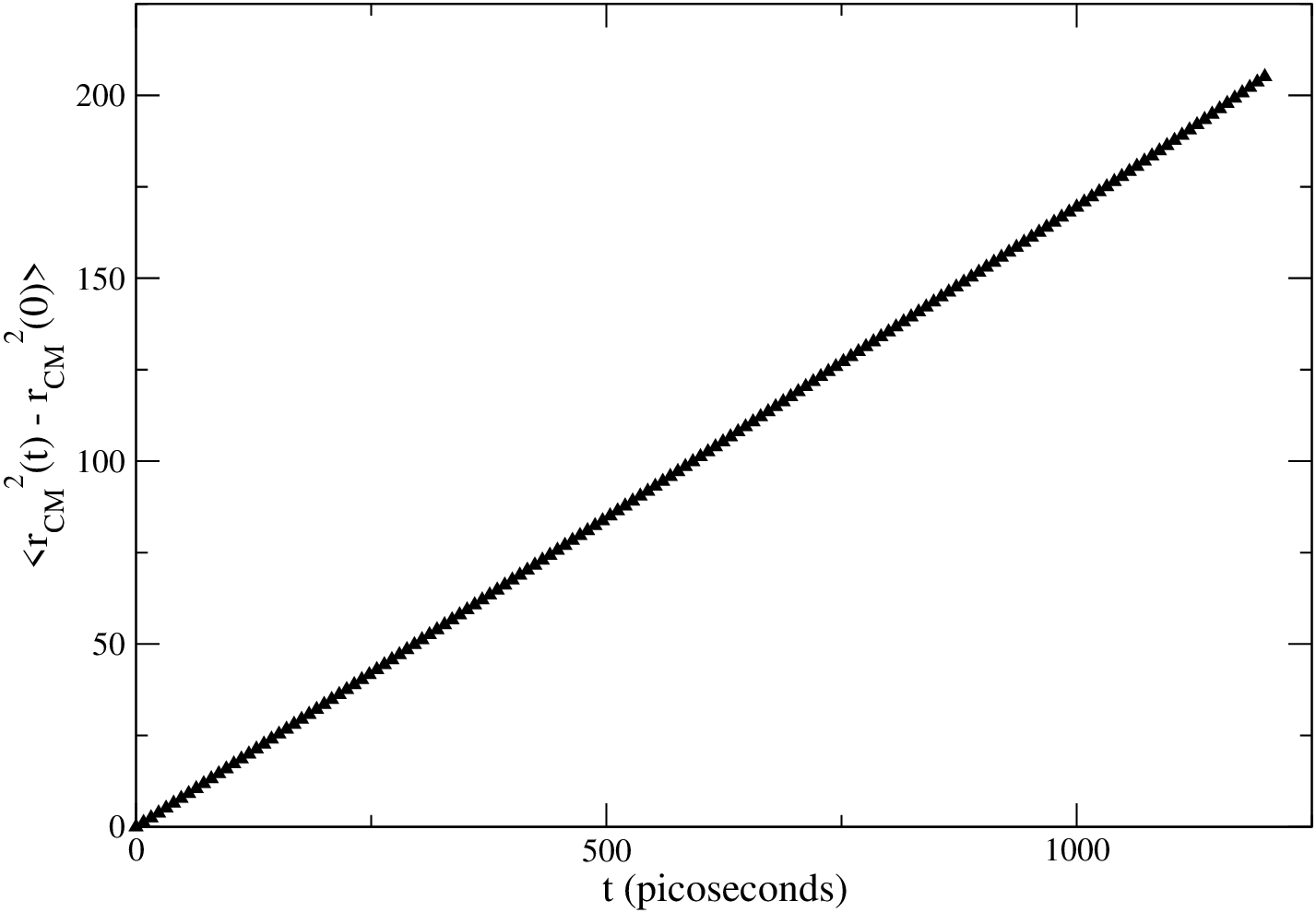}}}
\caption{}
\end{center}
\end{figure}


\bigskip

\begin{figure}
\begin{center}
\rotatebox{90}{\scalebox{0.8}[0.8]{\includegraphics{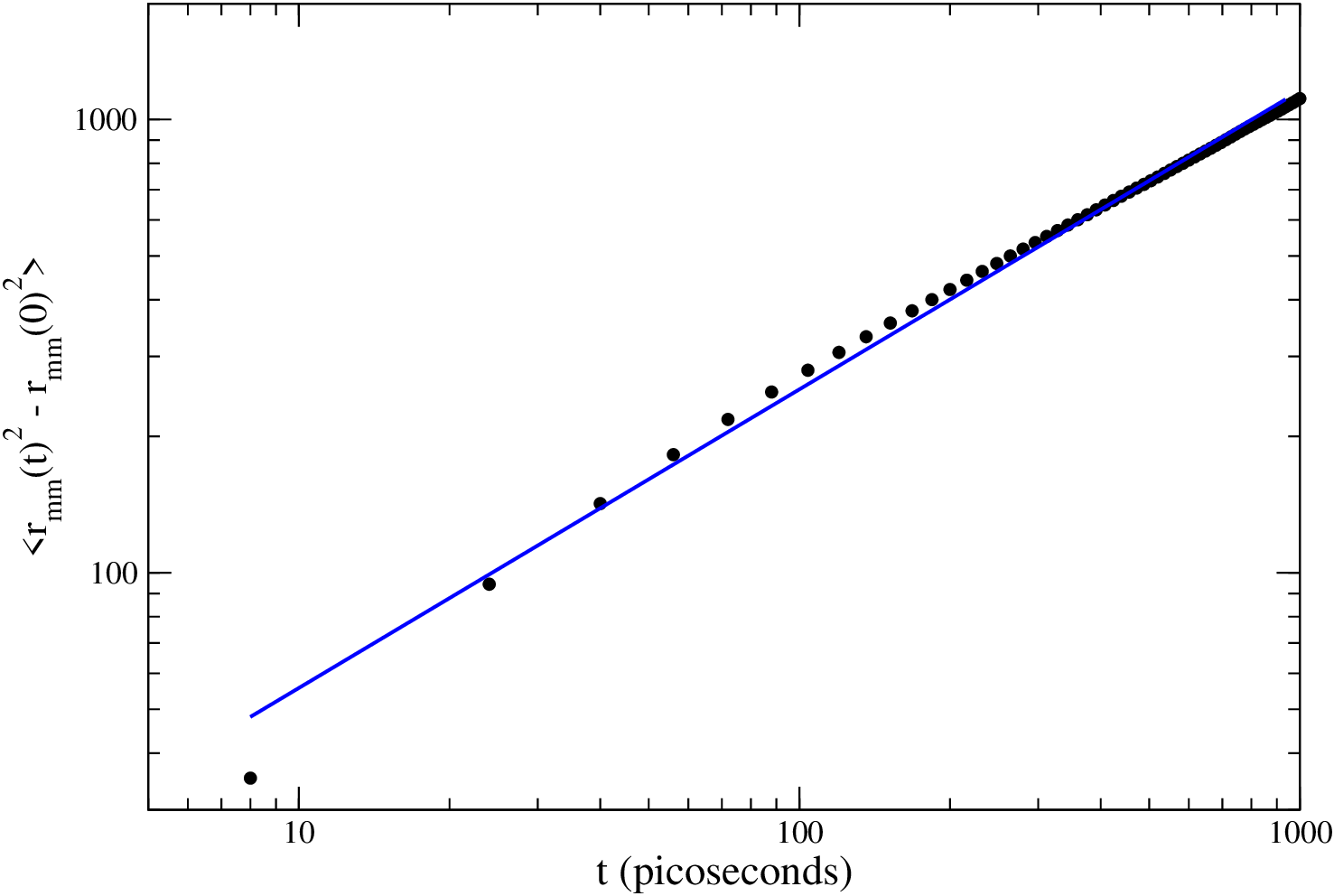}}}
\caption{}
\end{center}
\end{figure}


\bigskip

\begin{figure}
\begin{center}
\rotatebox{90}{\scalebox{0.8}[0.8]{\includegraphics{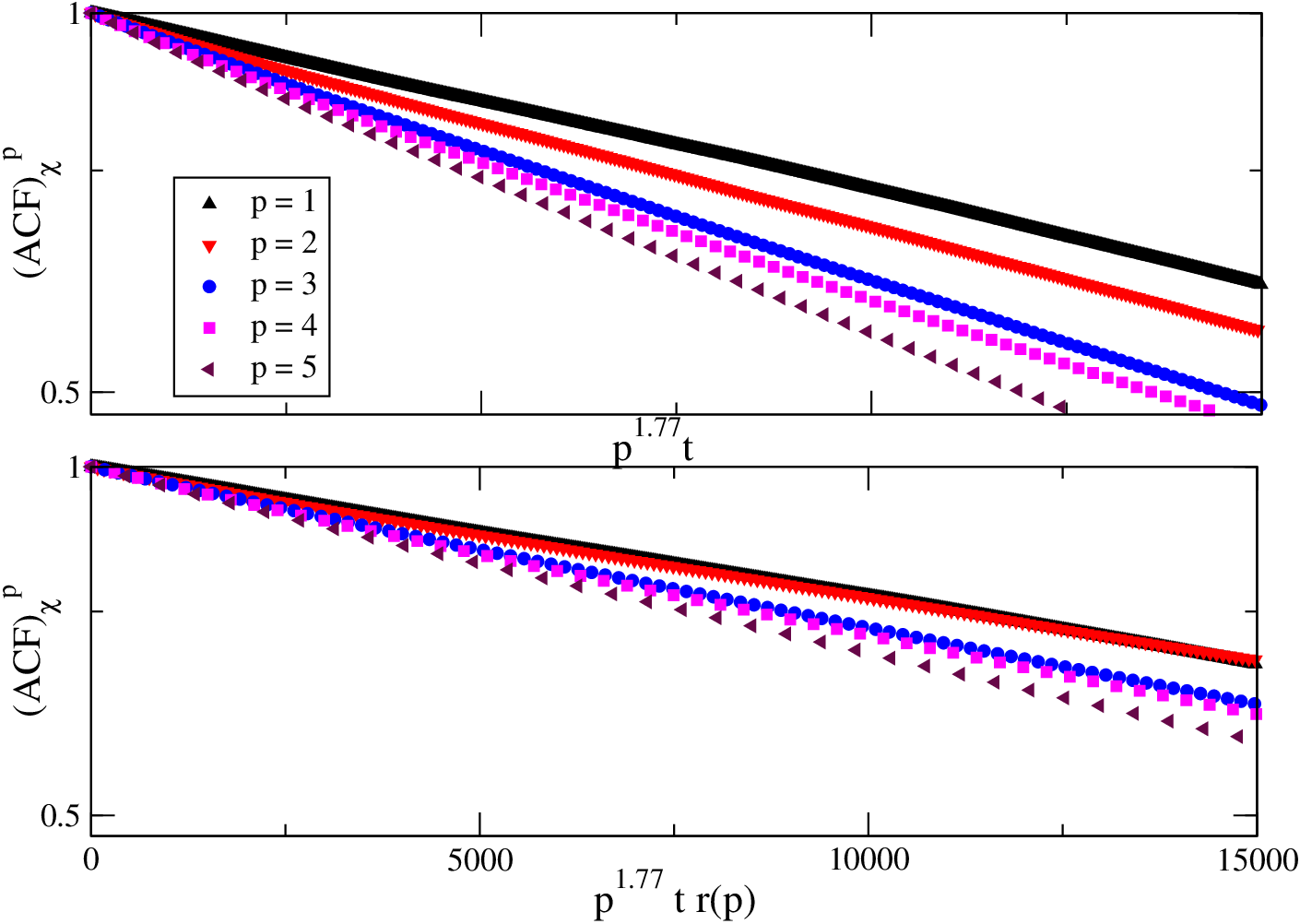}}}
\caption{}
\end{center}
\end{figure}


\bigskip

\begin{figure}
\begin{center}
\rotatebox{90}{\scalebox{0.8}[0.8]{\includegraphics{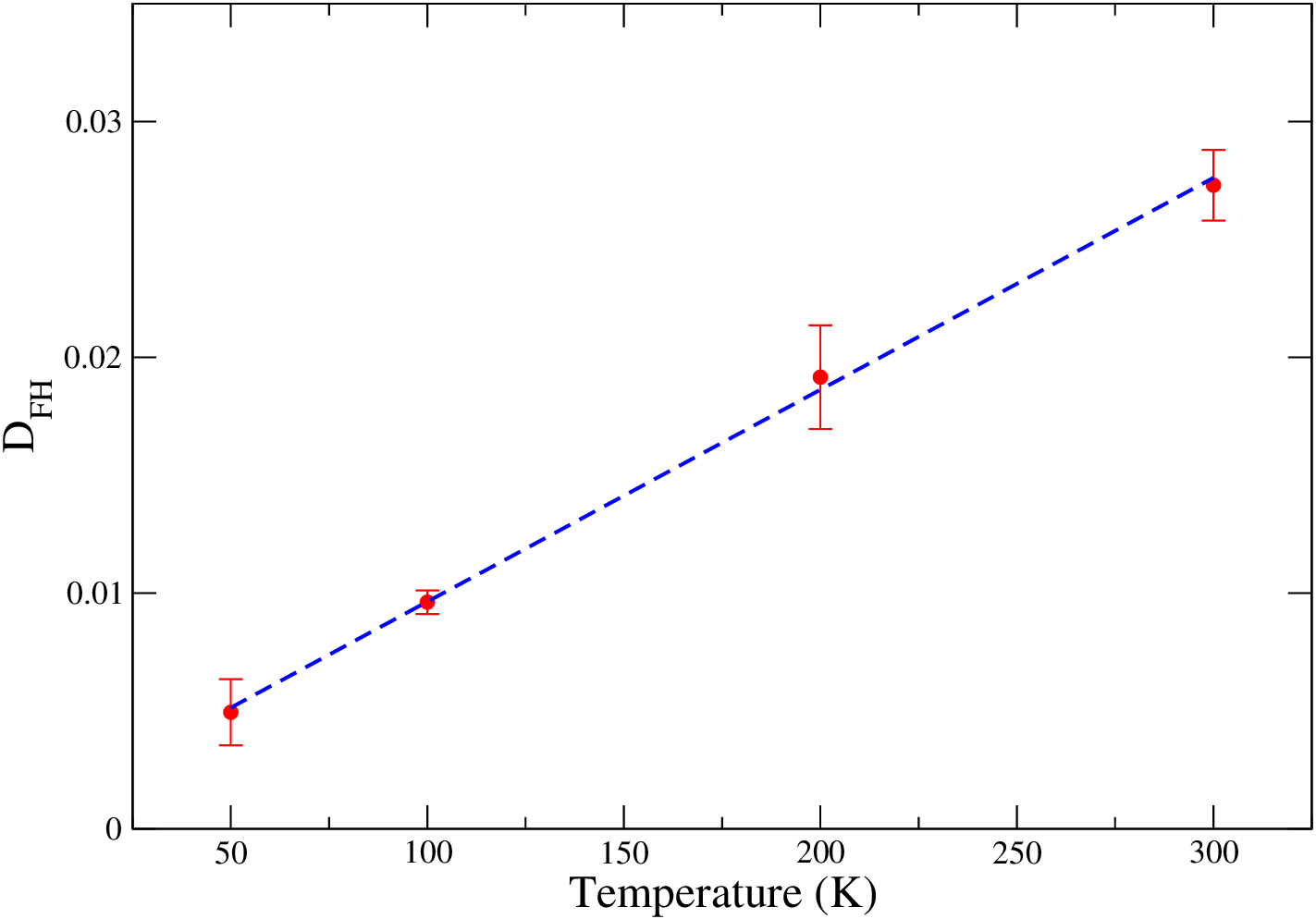}}}
\end{center}
\end{figure}


\bigskip

\begin{figure}
\begin{center}
\rotatebox{90}{\scalebox{0.8}[0.8]{\includegraphics{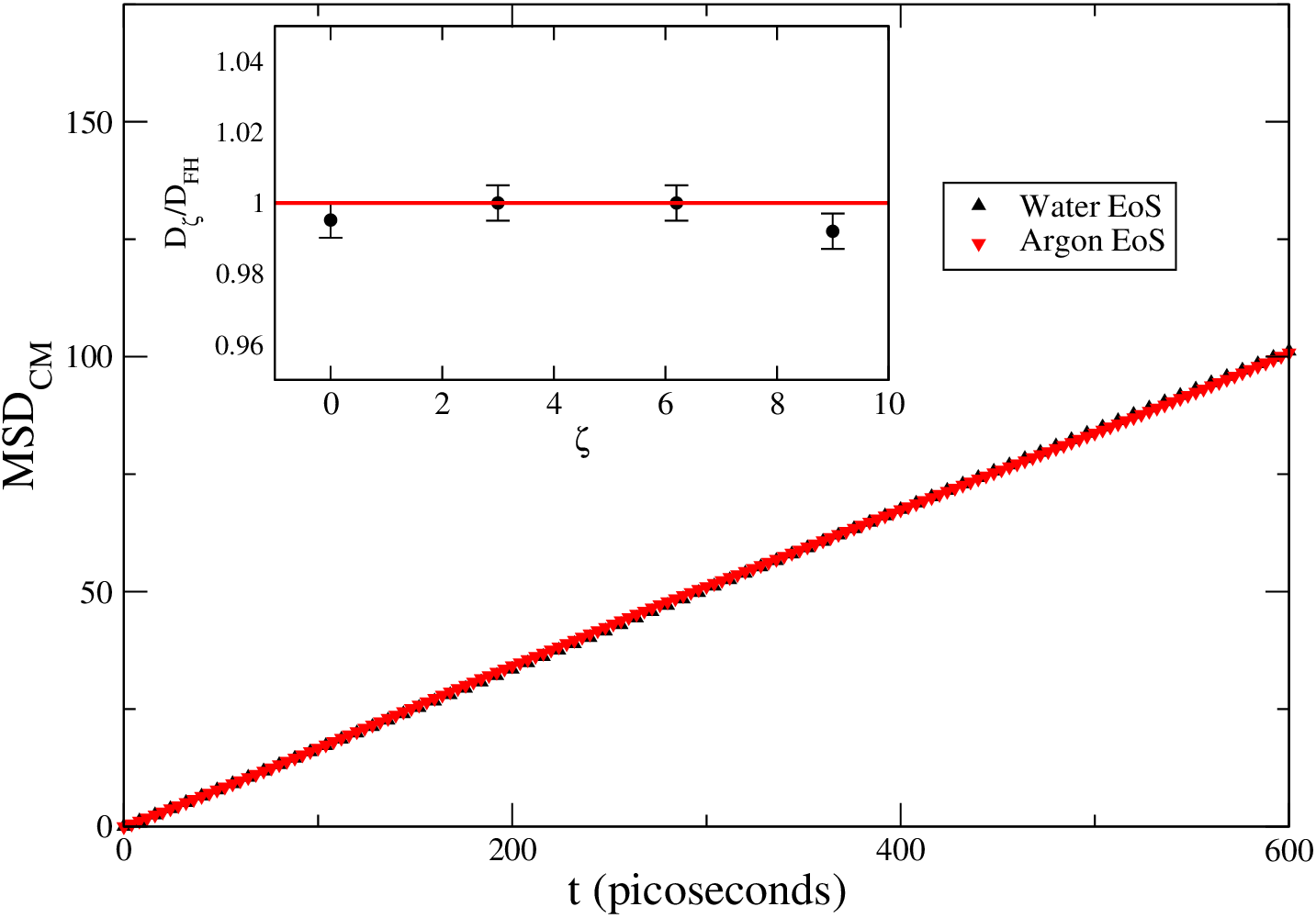}}}
\caption{}
\end{center}
\end{figure}


\bigskip

\begin{figure}
\begin{center}
\rotatebox{90}{\scalebox{0.8}[0.8]{\includegraphics{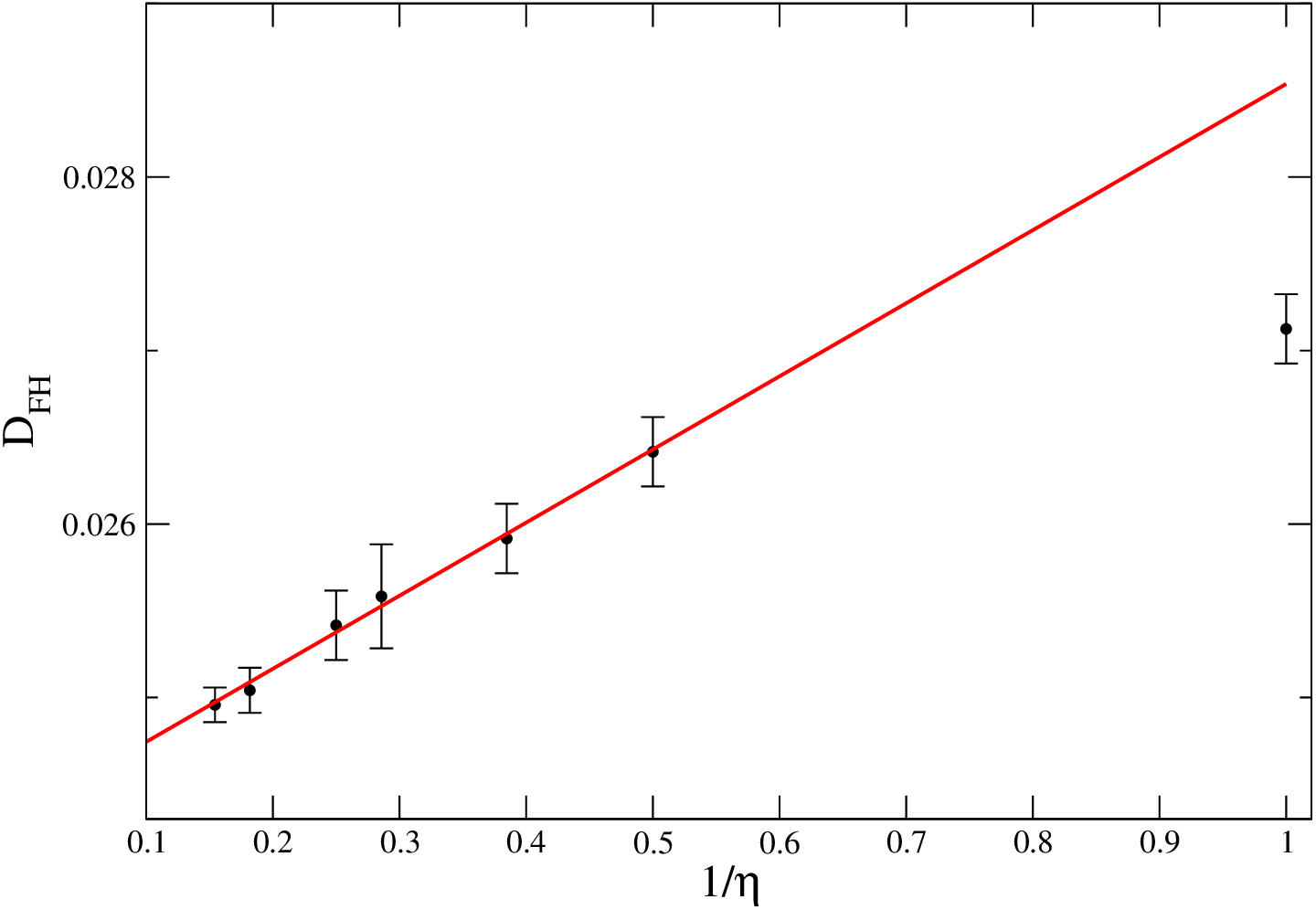}}}
\caption{}
\end{center}
\end{figure}

\bigskip

\begin{figure}
\begin{center}
\rotatebox{90}{\scalebox{0.8}[0.8]{\includegraphics{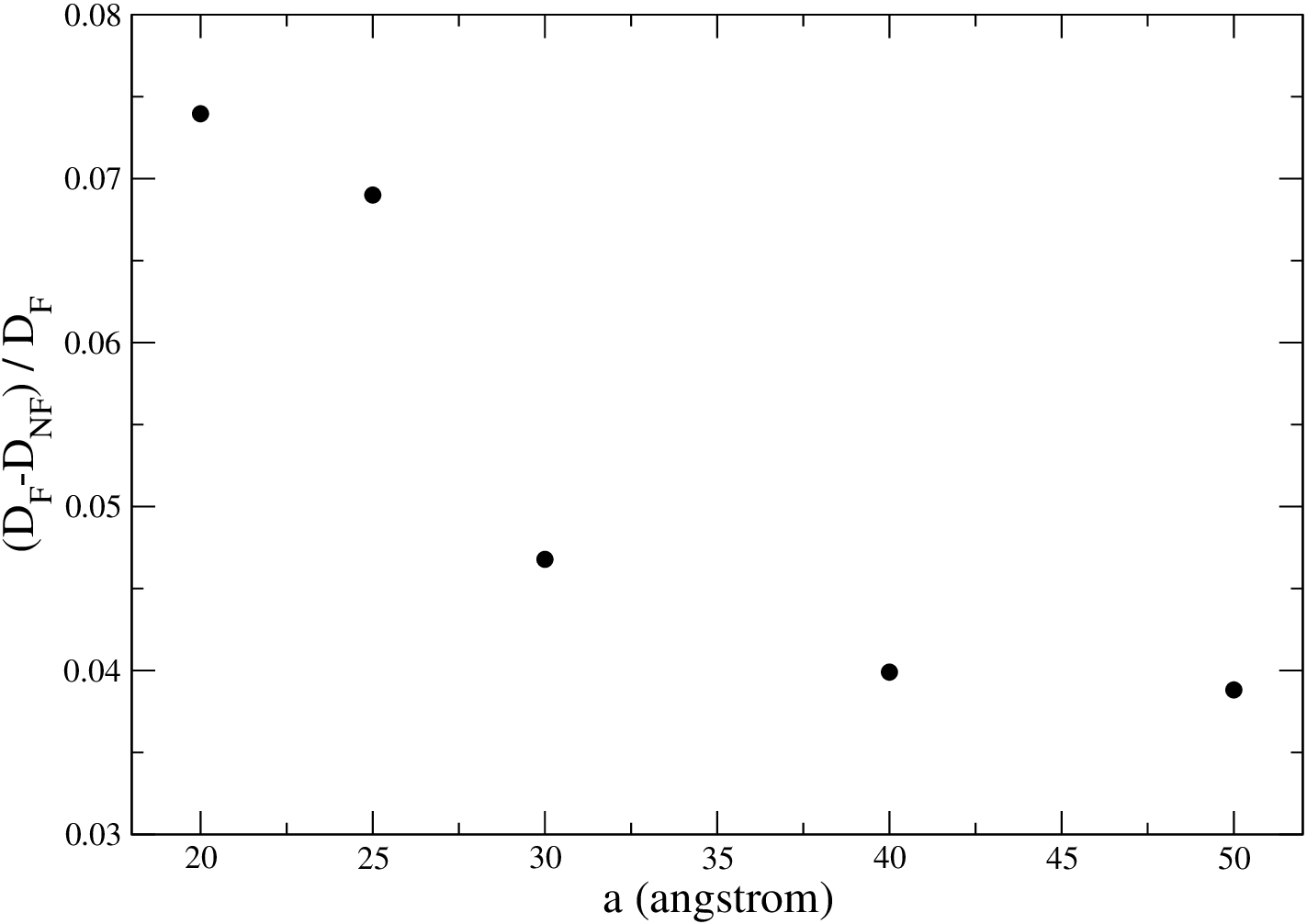}}}
\caption{}
\end{center}
\end{figure}


\end{document}